\documentclass[10pt,letterpaper,conference]{IEEEtran}
\IEEEoverridecommandlockouts

\usepackage{cite}
\usepackage{amsmath,amssymb,amsfonts}
\usepackage{algorithmic}
\usepackage{graphicx}
\usepackage{textcomp}
\usepackage{xcolor}
\usepackage{hyperref}
\usepackage{url}
\usepackage{graphicx}
\usepackage{multirow}
\usepackage{booktabs}
\usepackage{amsmath}
\usepackage{amssymb}
\usepackage{mathtools}
\usepackage{subcaption}
\usepackage{caption}
\usepackage{xcolor, colortbl}
\usepackage{cite}
\usepackage{comment}
\usepackage{threeparttable}
\usepackage{arydshln}
\IEEEoverridecommandlockouts 

\IEEEpubid{\begin{minipage}{\textwidth}\ \\[8pt] \centering
  \copyright~2025 IEEE. Personal use of this material is permitted. Permission from IEEE must be obtained for all other uses, in any current or future media, including reprinting/republishing this material for advertising or promotional purposes, creating new collective works, for resale or redistribution to servers or lists, or reuse of any copyrighted component of this work in other works.
\end{minipage}}

\begin{document}

\title{Unifying Diarization, Separation, and ASR with Multi-Speaker Encoder}
\author{
    \IEEEauthorblockN
        {Muhammad Shakeel\IEEEauthorrefmark{1},
        Yui Sudo\IEEEauthorrefmark{1},
        Yifan Peng\IEEEauthorrefmark{2},
        Chyi-Jiunn Lin\IEEEauthorrefmark{2},
        Shinji Watanabe\IEEEauthorrefmark{2}}
    \IEEEauthorblockA{\IEEEauthorrefmark{1}Honda Research Institute Japan, Japan\\}
    \IEEEauthorblockA{\IEEEauthorrefmark{2}Carnegie Mellon University, USA}
}

\maketitle
\IEEEpubidadjcol 

\begin{abstract}
    This paper presents a unified multi-speaker encoder (UME), a novel architecture that jointly learns representations for speaker diarization (SD), speech separation (SS), and multi-speaker automatic speech recognition (ASR) tasks using a shared speech foundational encoder. We leverage the hidden representations from multiple layers of UME as a residual weighted-sum encoding (RWSE) to effectively use information from different semantic levels, contributing to bottom-up alignment between tasks. This joint training approach captures the inherent interdependencies among the tasks, enhancing overall performance on overlapping speech data. Our evaluations demonstrate that UME substantially improves over the single-task baselines dedicated to SD, SS, and multi-speaker ASR on LibriMix evaluation sets. Notably, for SD, UME outperforms the previous studies, achieving diarization error rates of 1.37\% and 2.29\% on Libri2Mix and Libri3Mix evaluation sets, respectively.
\end{abstract}


\section{Introduction}
\label{introduction}
Speaker diarization (SD), speech separation (SS), and multi-speaker automatic speech recognition (ASR) are tasks of great importance that aim to comprehend and answer the question “who spoke what and when,” with applications to transcribing meetings and interviews, among others. Previous studies in SD \cite{taejin2022,Horiguchi,pyannote2023}, SS \cite{deLiang2018,Luo,tf-gridnet}, and multi-speaker ASR \cite{YanminQian,SekiHiroshi,Chang,kanda20b_interspeech} have focused primarily on improving the quality of single-task models that operate independently on acoustic information to separate or label speaker segments and transcribe the text in a speech-processing system \cite{deshraj1,CSS,ShinjiCHiME6}. A key limitation of training tasks independently is that inter-dependencies cannot be leveraged.

Most existing frameworks \cite{ShinjiCHiME6,zheng22f_interspeech,pyannote2023a} address this limitation by unifying speech-processing architectures \cite{Boeddeker,kalda1}. These architectures consist of either a joint ASR/SD \cite{li23o_interspeech,mao20b_interspeech,Cornell}, SS/ASR \cite{kanda2022streaming,XuankaiChang,neumann20_interspeech}, or a SD/SS \cite{Neumann,bando24_interspeech} task following a fixed optimal order that can vary depending on the target scene scenario \cite{mitrofanov24_chime,polok24_chime,niu24_chime,kamo24_chime}. These different target scenes suggest that we solve these tasks jointly, independent of the order, so all these tasks can benefit from each other. Lately, there has been a shift towards employing pre-trained speech foundation models (SFMs) \cite{wav2vec2,HuBERT,WavLM,whisper,OWSM-CTC} in end-to-end (E2E) systems, which effectively learn useful representations for various speech processing tasks \cite{SUPERB}. However they do not work well on multi-speaker conversation recognition. Additionally, it has been demonstrated that different layers encode different types of information in SD and ASR tasks \cite{SUPERB,WavLM}. Preliminary observation from these studies shows that intermediate layers of the encoder extract a rich hierarchy of information, e.g., in WavLM large \cite{WavLM}, initial layers and last layers are more critical for SD and ASR tasks. Therefore, it makes sense to utilize multiple layers to jointly optimize all SD, SS, and ASR tasks effectively. The question, therefore, naturally arises: \textit{can we build a unified model that leverages all encoder layers to optimize performance across multiple tasks?}

Motivated by the potential of SFMs and E2E speech processing, we propose a unified multi-speaker encoder (UME), a novel E2E speech-processing framework. The proposed framework is generalizable to use any SFM, E2E SD, SS and multi-speaker ASR task and jointly optimizes all these tasks into a single network with multitask learning to minimize the error accumulation for a speech processing framework. Additionally, by extracting features from all the layers of the OWSM (open Whisper style speech model \cite{OWSMv3.1}) v3.1 encoder and using it as a residual weighted-sum encoding (RWSE), we can learn better hidden representations from the encoder layers. We hypothesize that RWSE introduces information exchange and better bottom-up alignment to all the tasks from different semantic levels. We argue that UME framework should provide a shared representation space for SD, SS and multi-speaker ASR tasks and preferably have strong generalizability and learnability. We conduct extensive experiments on different design choices of UME on Libri2Mix \& Libri3Mix \cite{librimix} datasets. Our key contributions are:
\begin{itemize}
  \item We propose a unified speech-processing framework to jointly optimize the performance of SD, SS, and multi-speaker ASR tasks with hidden representations of the SFM encoder.
  \item We propose using RWSE of the pre-trained SFM encoder layers to unify and optimize the performance across diverse speech processing tasks.
  \item We demonstrate the effectiveness of our framework on two-speaker and three-speaker overlapped speech and obtain substantial performance improvement in each diarization, separation, and multi-speaker ASR task.
\end{itemize}

\section{Unified Multi-Speaker Encoder (UME)}
\label{sec:speech_tower}
\begin{figure}[t]
\begin{center}
{\resizebox{8cm}{!}{\includegraphics[width=\textwidth]{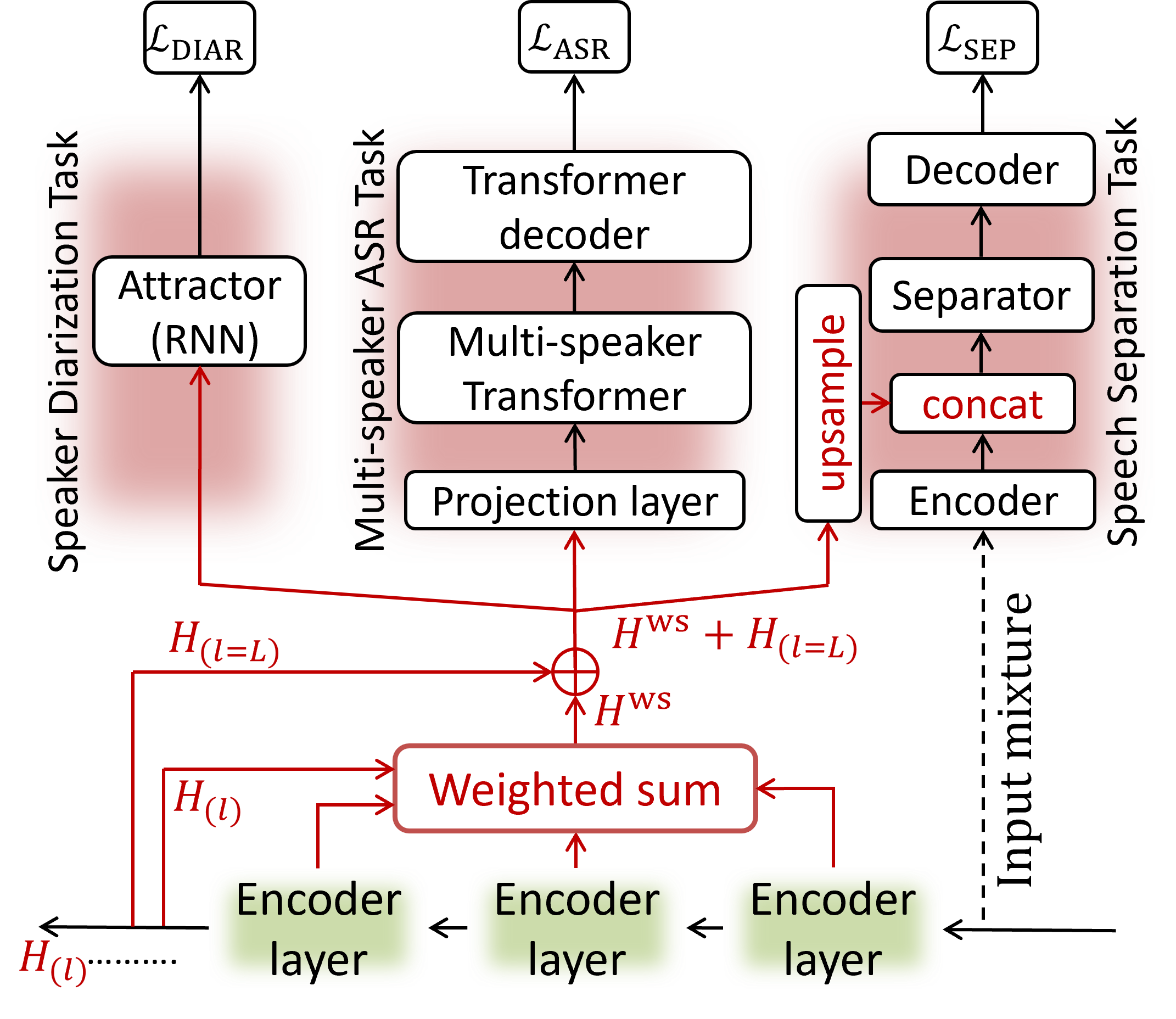}}}
\caption{Illustration of UME framework.}
\label{fig:UME} 
\end{center}
\vspace{-5px}
\end{figure}

\clearpage
Figure \ref{fig:UME} shows the overall framework of UME. It leverages the hidden representations through an RWSE of intermediate layers, which act as a bridge between SD, SS, and multi-speaker ASR tasks. This enables a comprehensive and detailed interaction from each layer of the SFM encoder. Note that our goal is not to develop a new encoder or speech processing tasks; in principle, one can apply any SFM encoder, SD, SS, or multi-speaker ASR tasks. 

We start with the $T$-length single-channel input speech mixture. \begin{math} X = \{x_{t} \in \mathbb{R}|t = 1,\cdots,T\} \end{math} of $C$ speakers. We define the input speech mixture in an anechoic condition by:
\begin{equation}
X = \sum_{c=1}^{C} Y^{c} \odot S^{c} + N,
\label{eq:inputspeechmixture}
\end{equation}
where, $S^{c} = \{s^{c}_{t} \in \mathbb{R} | t =1,...,T\}$ is the clean speech signal of speaker $c$, $Y^{c} = \{y^{c}_{t} \in [0,1] \mid t = 1,\cdots,T\}$ is a binary speech activity sequence indicating whether speaker $c$ is talking at time $t$ or not, and $N = \{n_{t} \in \mathbb{R} \mid t = 1,\cdots,T\}$ is the additive noise signal. Together, the activity sequences \( \{Y^c\}_{c=1}^C \) form the ground truth speaker label for the speaker diarization (SD) task described in Section~\ref{subsec:sd}.

\subsection{Speech Foundation Model Encoder}
\label{subsec:sfme}
We selected OWSMv3.1 \cite{OWSMv3.1} as the shared encoder due to its widespread recognition, reproducibility, open-source availability, and fast, efficient encoding capabilities. We can note that OWSM was trained on single-speaker speech-to-text tasks (i.e., no speaker tasks in pre-training). However, we can still adapt it to our multi-speaker setup. The speech encoder is a stack of $L$ E-Branchformer \cite{e-branchformer} encoder layers that transforms the input speech mixture $X$ of $C$ speakers into a $D^{\text{enc}}$-dimensional subsampled \begin{math}T^{\text{enc}}(<T)\end{math}-length hidden state representations \begin{math} H_{(l)} = \{\mathbf{h}_{(t,l)} \in \mathbb{R}^{D^{\text{enc}}}|t = 1,\cdots,T^{\text{enc}}\} \end{math}, where $l$ is a layer index from 1 to $L$. The simplified speech encoder is given by:
\begin{equation}
H_{(l)}= \mathrm{SpeechEnc}_{(l)}(X).
\label{eq:speechencoder}
\end{equation}
To integrate into task-specific models, all encoder layers are combined into a single feature vector via a weighted sum \cite{SUPERB}:  
\begin{equation}
H^{\text{ws}} = \sum_{l=1}^{L} \omega_{(l)}^{\text{task}} H_{(l)},
\label{eq:weighted_sum}
\end{equation}
where \begin{math} \omega^{\text{task}}_{(l)} \end{math} are softmax-normalized learnable weights that scale the representations from different encoder layers \textit{depending on each task}. Finally, RWSE is introduced by adding \( H^{\text{ws}} \) to the last encoder layer ($l=L$) to amplify the influence of the final layer according to \cite{WavLM} across tasks during training.
\begin{equation}
H^{\text{enc}} = H^{\text{ws}} + H_{(l=L)}.
\label{eq:rwse}
\end{equation}
\subsection{Speaker Diarization Task}
\label{subsec:sd}
Given the robust performance of E2E Neural Diarization (EEND) \cite{Horiguchi} with permutation invariant training (PIT) in estimating multi-speaker activities within an E2E framework, we adopt EEND for the SD task in the proposed UME E2E framework. SD involves predicting speaker activity as binary multi-class labels by estimating the speaker label sequence. We encode the hidden state representations $H^{\text{enc}}$ from the speech encoder and map the speaker activity probabilities $P^{c} (H^{\text{enc}}) \in \{ 0,1\}^{C}$ for speaker $c$ using a linear layer. We train EEND with PIT using speaker activity probabilities and the target speaker activity labels. We optimize the binary cross entropy-based (BCE) diarization loss ($\mathcal{L}_\text{diar}$) below:
\begin{equation}
\mathcal{L}_{\text{diar}} = \min_{\pi \in \mathcal{P}} \sum_{c=1}^{C} \mathrm{BCE}(Y^{\pi(c)}, P^{c}(H^{\text{enc}})),
\label{eq:loss_diar}
\end{equation}
where $\pi(c)$ is the mapping of $c$-th element under permutation $\pi$ and $\mathcal{P}$ is the set of all permutations over $c = \{1,\cdots,C\}$ and $Y^{\pi(c)}$ is the permuted reference of speaker labels.
\subsection{Speech Separation Task}
\label{subsec:ss}
We incorporate Conv-TasNet \cite{Luo} into our framework for its demonstrated adaptability in E2E-SS \cite{Neumann2020} task. While having suboptimal performance, its simple architecture makes it a well-known time-domain speech separation model capable of predicting speech signals using a fully convolutional encoder, separator, and decoder network. The input mixture is encoded via a 1-D convolutional encoder which is given by:
\begin{equation}
H^{\text{sep}} = \mathrm{ConvEnc}(X).
\label{eq:convencoder}
\end{equation}
We concatenate upsampled RWSE representations ($H^{\text{enc}}$) from Section \ref{subsec:sfme} with $H^{\text{sep}}$ and obtain $D^{\text{sep}}$-dimensional subsampled $T^{\text{sep}}(<T)$-length hidden state representations \begin{math} H^\text{concat} = \{\mathbf{h}_{t}^\text{concat} \in \mathbb{R}^{D^{\text{sep}}}\mid t = 1,\cdots,T^{\text{sep}}\} \end{math} which is given by: 
\begin{equation}
H^{\text{concat}} = \mathrm{concat}(H^{\text{sep}},\mathrm{upsample}(H^{\text{enc}})).
\label{eq:convencoder_concat}
\end{equation}
These representations are then processed using stacked 1-D dilated temporal convolutional networks (TCNs) to extract the embedding sequence \begin{math} E = \{\mathbf{e}_{t} \in \mathbb{R}^{K}\mid t = 1,\cdots,T^{\text{sep}}\} \end{math} in \eqref{eq:tcn}:
\begin{equation}
E = \mathrm{TCN}(\mathrm{Conv_{1\times1}}(\mathrm{LayerNorm}(H^{\text{concat}}))).
\label{eq:tcn}
\end{equation}
The separation network then estimates a mask sequence $M^{c} = \{\mathbf{m}^{c}_{t}\in [0,1]^O |t = 1,\cdots,T^{\text{sep}}\}$ in \eqref{eq:masks} for each source: 
\begin{equation}
M^c = \sigma([\mathrm{Conv_{1\times1}}(\mathrm{PReLU}(E))]_{c}),
\label{eq:masks}
\end{equation}
where it computes the source-specific representation sequence $D^c = \{ \mathbf{d}^{c}_{t} \in \mathbb{R}^J \mid t = 1, \dots, T^{\text{sep}} \}$ by applying the masks using element-wise multiplication $\odot$ in \eqref{eq:elementwise}.
\begin{equation}
D^c = H^{\text{concat}} \odot M^c.
\label{eq:elementwise}
\end{equation}
Finally, a 1-D transposed convolutional decoder reconstructs the time-domain waveform for each source $\hat{S}^{c} = \{\hat{s}_t^c \in \mathbb{R} \mid t = 1, \dots, T\}$ in \eqref{eq:sepdecoder} and optimizes the $\mathrm{SI-SDR}$ loss ($\mathcal{L}_{\text{sep}}$) in \eqref{eq:si-sdr}.
\begin{equation}
\hat{S}^{c} = \mathrm{Decoder}(D^{c}),
\label{eq:sepdecoder}
\end{equation}
\begin{equation}
\mathcal{L}_{\text{sep}} = \min_{\pi \in \mathcal{P}} \left( 
-10 \sum_{c=1}^{C} \log_{10} \left( 
\frac{\left\lVert \frac{\langle \hat{S}^{c}, S^{\pi(c)} \rangle S^{\pi(c)}}{\left\lVert S^{\pi(c)} \right\rVert^2} \right\rVert^2}
{\left\lVert \hat{S}^{c} - \frac{\langle \hat{S}^{c}, S^{\pi(c)} \rangle S^{\pi(c)}}{\left\lVert S^{\pi(c)} \right\rVert^2} \right\rVert^2}
\right) 
\right),
\label{eq:si-sdr}
\end{equation}
\subsection{Multi-speaker ASR Task}
\label{subsec:multiasr}
The multi-speaker ASR task, as adopted from \cite{Chang}, extends a joint connectionist temporal classification (CTC)/attention-based framework to recognize speech from multiple speakers. We encode the input hidden state representations (see Section \ref{subsec:sfme}) from the speech encoder. Subsequently, each speaker’s speech is extracted through $C$ speaker-differentiating encoder blocks (SpeakerEnc$_{\text{SD}}^{c}$). These speaker-dependent features are then transformed into $D^{\text{asr}}$-dimensional subsampled \begin{math}T^{\text{asr}}(<T^{\text{enc}})\end{math}-length hidden state representations \begin{math} H^{\text{asr}}_{c} = \{\mathbf{h}^{\text{asr}}_{(t,c)} \in \mathbb{R}^{D^{\text{asr}}}\mid t = 1,\cdots,T^{\text{asr}}\} \end{math}, where \begin{math} c = \{1,\cdots,C\}\end{math} enumerates speakers.
\begin{equation}
H^{\text{asr}}_{c} = \mathrm{SpeakerEnc}_{\text{SD}}^{c}(H^{\text{enc}}).
\label{eq:speakerencoder}
\end{equation}
The attention-based decoder generates a speaker-specific $U$-length output token sequence \begin{math} W^{c} = \{w^{c}_{u} \in \mathcal{V}\mid u = 1,\cdots,U\} \end{math}, where $w^{c}_{u}$ is an output token at position $u$ in the vocabulary $\mathcal{V}$ for speakers $c = 1,\cdots,C$. Specifically, PIT (see Section \ref{subsec:sd}) is employed to control the reference sequences $W^c$ permutation $\pi$ using the CTC loss ($\mathcal{L}_{\text{ctc}}$) immediately after the encoder. Finally, the loss for the multi-speaker ASR ($\mathcal{L}_{\text{asr}}$) task is optimized using CTC and cross-entropy loss of the attention decoder ($\mathcal{L}_{\text{att}}$):
\begin{align}
\mathcal{L}_{\text{asr}} = \min_{\pi \in \mathcal{P}} \sum_{c=1}^{C} \Big[
& \lambda_{\text{ctc}} \, \mathcal{L}_{\text{ctc}}(W^{\pi(c)}, H^{\text{asr}}_{c}) \notag \\
& + (1 - \lambda_{\text{ctc}}) \, \mathcal{L}_{\text{att}}(W^{\pi(c)}, H^{\text{asr}}_{c})
\Big],
\label{eq:loss_asr}
\end{align}

We optimize the final loss function as a weighted sum of $\mathcal{L}_{\text{diar}}$ in \eqref{eq:loss_diar}, $\mathcal{L}_{\text{sep}}$ in \eqref{eq:si-sdr} and $\mathcal{L}_{\text{asr}}$ in \eqref{eq:loss_asr}. $\lambda_{\text{diar}}$, $\lambda_{\text{sep}}$, and $\lambda_{\text{asr}}$ are the weighting hyperparameters which are optimized empirically.
\begin{equation}
\mathcal{L}_{\text{all}}= \lambda_{\text{diar}}\mathcal{L}_{\text{diar}} + \lambda_{\text{sep}}\mathcal{L}_{\text{sep}} + \lambda_{\text{asr}}\mathcal{L}_{\text{asr}}.
\label{eq:loss_all}
\end{equation}
Thus, we can integrate all diarization, separation, and ASR tasks in a unified multi-speaker-encoder architecture.

\section{Experiments}
\label{sec:experiments}

\subsection{Dataset and evaluation metrics}
\label{subsec:dataset}

In UME, we aim to optimize all three tasks: diarization, separation, and multi-speaker ASR, using multi-task learning in a unified framework. We require three ground truths to objectively evaluate performance, i.e., diarization labels, separated sources, and text for each speaker. While real-world multiparty datasets \cite{AMI,hitachijhudihardiiisystem} exist for diarization-only tasks, they often need separated sources and the number of speakers to be known for ASR. Therefore, we employ simulated conversation-like open-source datasets for training and evaluation of all three tasks. For training, we use LibriMix, which combines LibriSpeech samples with and without WHAM! noise, supporting two-speaker (Libri2Mix) and three-speaker (Libri3Mix) mixtures. We adopt a 16kHz sampling rate, the ``mixboth” and  ``mixclean" method with 100\% overlap. Moreover, we selected the ``max'' mode in LibriMix, as the ASR task is unfeasible in the ``min'' mode due to the truncation of speech signals on minimum-length sequences. We evaluate our framework using the Libri2Mix and Libri3Mix datasets \cite{librimix} with a complete 100\% overlap, as well as the LibriSpeech2Mix and LibriSpeech3Mix datasets \cite{kanda20b_interspeech}, which include a partial random overlap of at least 0.5 seconds.

We evaluate diarization using the diarization error rate (DER\%) \cite{DER} with a 0.0s, 0.25s collar tolerance and 11-frame median filtering. For separation, we report three metrics: STOI (dB) \cite{STOI}, SDR (dB) \cite{SDR} and SI-SNR (dB) \cite{Luo}. Multi-speaker ASR performance is measured using WER with the optimum permutation following the prior studies in \cite{Chang} and \cite{kanda20b_interspeech}. Unlike \cite{ShinjiCHiME6}, our WER computation is independent of the diarization branch i.e., diarization frame information is not used during the multi-speaker ASR decoding process.
\begin{table*}[!ht]
\caption{The results are presented using WER ($\downarrow$) for multi-speaker ASR on Libri2Mix (L2Mix), Libri3Mix (L3Mix), LibriSpeech2Mix (LS2Mix) and LibriSpeech3Mix (LS3Mix) evaluation sets. ($\lambda_{\text{asr}},\lambda_{\text{diar}}, \lambda_{\text{sep}}$) denote training weights next to UME. \textbf{Bold:} the best result on clean speech mixtures. \underline{Underlined:} our best result on noisy speech mixtures.}
\label{table:twospeaker}
\begin{center}
\resizebox {0.65\linewidth} {!} 
{
\begin{threeparttable}[b]
\begin{tabular}{@{}llllll}
    \toprule
\multicolumn{1}{l}{\bf ID}                                                         & \multicolumn{1}{l}{\hspace{2em} \bf Model}  & \multicolumn{1}{c}{\bf L2Mix\tnote{$\dagger$}}  & \multicolumn{1}{c}{\bf L3Mix\tnote{$\dagger$}} & \multicolumn{1}{c}{ \bf LS2Mix\tnote{$\ddagger$}} & \multicolumn{1}{c}{\bf LS3Mix\tnote{$\ddagger$}} \\
    \midrule
\rowcolor{gray!30}\multicolumn{6}{c}{\bf{\textit{Baselines}} $\rightarrow$  \textbf{\textit{Training set: LibriSpeechMix (960 hours, speech only, mode= partial overlap)}}} \\
\rowcolor{gray!10}\multicolumn{1}{l}{} &\hspace{2em}PIT LSTM-AED \cite{kanda20b_interspeech}             &  N/A                           &N/A                            & 11.9\tnote{$\divideontimes$}                  & 52.3\tnote{$\divideontimes$}   \\
\rowcolor{gray!10}\multicolumn{1}{l}{} &\hspace{2em}SOT \cite{kanda20b_interspeech}                      &  N/A                           &N/A                            & 11.2\tnote{$\divideontimes$}                  & 24.0\tnote{$\divideontimes$}    \\
\rowcolor{gray!10}\multicolumn{1}{l}{} &\hspace{2em}SURT \cite{lianglu}                                  &  N/A                           &N/A                            & 7.2\tnote{$\divideontimes$}                   & N/A    \\
\rowcolor{gray!10}\multicolumn{1}{l}{} &\hspace{2em}t-SOT \cite{kanda2022streaming}                      &  N/A                           &N/A                            & 5.8\tnote{$\divideontimes$}                   & N/A    \\
\rowcolor{gray!10}\multicolumn{1}{l}{} &\hspace{2em}SOT-Conformer \cite{kanda21b_interspeech}            &  N/A                           &N/A                            & 4.9\tnote{$\divideontimes$}                   & \textbf{6.2}\tnote{$\divideontimes$}    \\
\rowcolor{gray!10}\multicolumn{1}{l}{} &\hspace{2em}Whisper-medium-SS-TTI \cite{meng24c_interspeech}     &  N/A                           & N/A                           & \textbf{4.0}\tnote{$\divideontimes$}          & 7.5\tnote{$\divideontimes$}    \\
 \midrule
\rowcolor{brown!30}\multicolumn{6}{l}{\bf{\textit{Baseline}} $\rightarrow$ \textbf{\textit{Training set: LibriMix (460 hours, mixboth: speech $+$ noise, mode= max)}}} \\
\rowcolor{brown!10}\multicolumn{1}{l}{} & \hspace{2em}Multi-speaker AED \cite{Chang} (reproduced)        & 24.4                           &N/A                            & 12.7                                          & N/A      \\
\rowcolor{pink!100}\multicolumn{6}{l}{\textbf{\textit{Proposed}} $\rightarrow$ \textbf{\textit{Training set: LibriMix (460 hours, mixboth: speech $+$ noise, mode= max)}}} \\
\rowcolor{pink!70}\multicolumn{1}{l}{\textbf{A1}} & \hspace{1em} \textbf{\textit{w/o weighted sum}}      &                                &                          &                               &                        \\
\rowcolor{pink!30}\multicolumn{1}{l}{} & \hspace{2em} UME ($\lambda_{\text{asr}} = 1.0$)                 & 25.0                           & \underline{26.4}         & 13.0                          & 16.0                   \\
\rowcolor{pink!30}\multicolumn{1}{l}{} & \hspace{2em} UME (0.33, 0.33, 0.34)                             & 22.7                           & div.\tnote{$\S$}         & 11.0                          & div.\tnote{$\S$}       \\
\rowcolor{pink!30}\multicolumn{1}{l}{} & \hspace{3em} + ASR init.                                        & 21.1                           & 26.5                     & \underline{9.2}               & \underline{15.7}       \\
\rowcolor{pink!30}\multicolumn{1}{l}{} & \hspace{2em} UME (0.1, 0.1, 0.8)                                & 22.4                           & div.\tnote{$\S$}         & 11.9                          & div.\tnote{$\S$}       \\
\rowcolor{pink!30}\multicolumn{1}{l}{} & \hspace{3em} + ASR init.                                        & N/A                            & 27.3                     &  N/A                          & 20.3                   \\
\rowcolor{pink!70}\multicolumn{1}{l}{\textbf{A2}} & \hspace{1em}  \textbf{\textit{w/ weighted sum}}      &                                &                          &                               &                        \\
\rowcolor{pink!30}\multicolumn{1}{l}{} & \hspace{2em} UME (0.1, 0.1, 0.8)                                & 25.5                           & div.\tnote{$\S$}         & 12.8                          & div.\tnote{$\S$}       \\
\rowcolor{pink!70}\multicolumn{1}{l}{\textbf{A3}} & \hspace{1em}  \textbf{\textit{w/ RWSE}}              &                                &                          &                               &                        \\
\rowcolor{pink!30}\multicolumn{1}{l}{} & \hspace{2em} UME ASR init. (0.33, 0.33, 0.34)                   & \underline{19.6}               & 27.1                     & 10.3                          & 18.0                   \\
 \midrule
\rowcolor{brown!30}\multicolumn{6}{l}{\bf{\textit{Baseline}} $\rightarrow$ \textbf{\textit{Training set: LibriMix (unk. hours, mixclean: speech only, mode= max)}}} \\
\rowcolor{brown!10}\multicolumn{1}{l}{} &\hspace{2em}Whisper-medium-SS-TTI \cite{meng24c_interspeech}    &  6.56                          & 21.47                    & N/A                           & N/A                    \\
\rowcolor{pink!100}\multicolumn{6}{l}{\bf{\textit{Proposed}} $\rightarrow$ \textbf{\textit{Training set: LibriMix (460 hours, mixclean: speech only, mode= max)}}} \\
\rowcolor{pink!70}\multicolumn{1}{l}{\textbf{A4}} & \hspace{1em}  \textbf{\textit{w/ RWSE}}              &                                &                          &                               &                        \\
\rowcolor{pink!30}\multicolumn{1}{l}{} &\hspace{2em} UME ASR init. (0.33, 0.33, 0.34)                    & \textbf{6.4}                   & \textbf{15.9}            & 5.5                           & 12.9                   \\
  \bottomrule
\end{tabular}
     \begin{tablenotes}
       \item \footnotesize{$\dagger$ The test set matches the training set (mixboth $\rightarrow$ mixboth, mixclean $\rightarrow$ mixclean).}
       \item \footnotesize{$\ddagger$ The test set contains only clean speech mixtures with partial overlap.}
       \item \footnotesize{$\divideontimes$ The test set matches the training set.}
       \item \footnotesize{$\S$ For three-speaker case, training diverged (div.) w/o ASR initialization.}
     \end{tablenotes}
\end{threeparttable}
}
\vspace{-20px}
\end{center}
\end{table*}
\subsection{Implementation details}
\label{subsec:implementation}
UME employs the pre-trained supervised SFM encoder OWSMv3.1 \cite{OWSMv3.1} (medium) as a shared feature extractor for all tasks. We use learnable weights with RWSE (see Section \ref{subsec:sfme}) to optimize all OWSMv3.1 layers jointly. For diarization following the hyperparameters reported in \cite{Horiguchi} (see Section \ref{subsec:sd}), we input the RWSE features (frame length: 400, frameshift: 640 samples) into a 1-layer RNN attractor (hidden size: 1024). For separation (see Section \ref{subsec:ss}), we concatenate 1024-dimensional OWSMv3.1 hidden representations with 256-dimensional Conv-TasNet \cite{Luo} encoded features. Since OWSMv3.1 has a 40 ms frameshift, we up-sample its features for time alignment, as in \eqref{eq:convencoder_concat} before feeding them into three TCN blocks with eight convolutional layers (hidden size: 512). For multi-speaker ASR (see Section \ref{subsec:multiasr}), we project OWSMv3.1 features to 128 dimensions, then process them with a post-encoder having four SpeakerEnc$_{\text{SD}}^{c}$ transformer blocks (2048 linear units, 256 input dimension) and a convolutional layer with a subsampling factor of four. It is jointly optimized ($\lambda_{\text{ctc}}$=0.2 \& $\lambda_{\text{att}}$=0.8) with six transformer-based decoder blocks (2048 linear units, 256 input dimension) and a CTC layer following \cite{Chang}. During training in UME, we initialize the encoder parameters with the pre-trained OWSMv3.1 medium encoder and fine-tune the encoder layers for 70 epochs, while all task-specific parameters have a flat start (i.e., no parameter initialization for task-specific layers) and are trained for 70 epochs. In the ASR-initialized UME version, the ASR model is pre-trained separately for 30 epochs before integration. We use the AdamW optimizer with an initial learning rate of $4\cdot10^{-4}$ (empirically tuned) and weight decay of $1\cdot10^{-6}$ warmed up for 10,000 steps. Training on four A100 80GB GPUs took six days, with an average batch size of 44, dynamically adjusted using the ESPnet \cite{espnet} numel batch type. For task-specific weighting, we use an equal-weight scalarization approach \cite{Bazgan2022}, assigning $\lambda_{\text{asr}}$=0.33, $\lambda_{\text{diar}}$=0.33, $\lambda_{\text{sep}}$=0.34 to balance tasks. This approach assumes that the tasks are cooperative rather than conflicting. While we also explored a two-stage weight optimization strategy inspired by \cite{sudo25_taslp}, however, it led to performance degradation. 
\begin{table*}[ht]
\caption{DERs (\%) for two-speaker and three-speaker evaluations. No collar tolerance was allowed. ($\lambda_{\text{asr}},\lambda_{\text{diar}}, \lambda_{\text{sep}}$) denote training weights next to UME. \textbf{Bold:} the proposed method outperforms the baseline. \textbf{\underline{Underlined:}} the best result.}
\label{table:der}
\begin{center}
\resizebox {0.60\linewidth} {!} {
\begin{threeparttable}[b]
\begin{tabular}{@{}lp{5cm}ll}
    \toprule
    \multicolumn{1}{l}{\bf ID}                         & \multicolumn{1}{l}{\hspace{2em} \bf Model}       & \multicolumn{1}{l}{\bf Libri2Mix}                 & \multicolumn{1}{l}{\bf Libri3Mix}\\
    \midrule
    \rowcolor{gray!30}\multicolumn{4}{l}{\bf{\textit{Baseline}}$\rightarrow$\textbf{\textit{Training set: LibriMix (460 hours, mixboth: speech $+$ noise, mode= max)}}} \\
    \rowcolor{gray!10}\multicolumn{1}{l}{}             & EEND \cite{Horiguchi} (reprod.)                  &  4.62                                             & N/A\\
    \midrule
    \rowcolor{pink!100}\multicolumn{4}{l}{\bf{\textit{Proposed}}$\rightarrow$\textbf{\textit{Training set: LibriMix (460 hours, mixboth: speech $+$ noise, mode= max)}}}    \\
    \rowcolor{pink!70}\multicolumn{1}{l}{\textbf{A1}}  & \textbf{\textit{w/o weighted sum}}               &                                                   & \\
    \rowcolor{pink!30}\multicolumn{1}{l}{}             & \hspace{1em} UME ($\lambda_{\text{diar}} = 1.0$) & \textbf{2.91}                                     & \textbf{3.26}\\
    \rowcolor{pink!30}\multicolumn{1}{l}{}             & \hspace{1em} UME (0.1, 0.1, 0.8)                 & \textbf{2.28}                                     & div.\tnote{$\S$}\\            
    \rowcolor{pink!70}\multicolumn{1}{l}{\textbf{A2}}  & \textbf{\textit{w/ weighted sum}}                &                                                   & \\
    \rowcolor{pink!30}\multicolumn{1}{l}{}             & \hspace{1em} UME (0.33, 0.33, 0.34)              & \textbf{2.26}                                     & div.\tnote{$\S$} \\
    \rowcolor{pink!30}\multicolumn{1}{l}{}             & \hspace{2em} + ASR init.                         & \textbf{2.45}                                     & \textbf{3.15}\\ 
    \rowcolor{pink!30}\multicolumn{1}{l}{}             & \hspace{1em} UME (0.1, 0.1, 0.8)                 & \textbf{2.19}                                     & div.\tnote{$\S$}\\
    \rowcolor{pink!30}\multicolumn{1}{l}{}             & \hspace{2em} + ASR init.                         & N/A                                               & \textbf{2.84}\\
    \rowcolor{pink!70}\multicolumn{1}{l}{\textbf{A3}}  & \textbf{\textit{w/ RWSE}}                        &                                                   & \\
    \rowcolor{pink!30}\multicolumn{1}{l}{}             & \hspace{1em} UME ASR init. (0.33, 0.33, 0.34)    & \underline{\textbf{2.14}}\tnote{$\dagger$} & \underline{\textbf{3.01}}\tnote{$\dagger$} \\
    \midrule
    \rowcolor{gray!30}\multicolumn{4}{l}{\bf{\textit{Baselines}}$\rightarrow$\textbf{\textit{Training set: LibriMix (100 hours, mixclean: speech only, mode= max)}}} \\
    \rowcolor{gray!10}\multicolumn{1}{l}{}             & \hspace{1em} HuBERT Large \cite{HuBERT}          & 5.75                                              & N/A \\
    \rowcolor{gray!10}\multicolumn{1}{l}{}             & \hspace{1em} wav2vec 2.0 Large \cite{wav2vec2}   & 5.62                                              & N/A \\
    \rowcolor{gray!10}\multicolumn{1}{l}{}             & \hspace{1em} WavLM Large \cite{WavLM}            & 3.24                                              & N/A\\
    \rowcolor{pink!100}\multicolumn{4}{l}{\bf{\textit{Proposed}}$\rightarrow$\textbf{\textit{Training set: LibriMix (460 hours, mixclean: speech only, mode= max)}}} \\
    \rowcolor{pink!70}\multicolumn{1}{l}{\textbf{A4}}  & \textbf{\textit{w/ RWSE}}                       &                                                   & \\
    \rowcolor{pink!30}\multicolumn{1}{l}{}             & \hspace{1em} UME ASR init. (0.33, 0.33, 0.34)   & \underline{\textbf{1.37}}\tnote{$\ast$}           & \underline{\textbf{2.29}}\tnote{$\ast$} \\
    \bottomrule
\end{tabular}
     \begin{tablenotes}
       \item \footnotesize{$\S$ For three-speaker case, training diverged (div.) w/o ASR initialization.}
       \item \footnotesize{$\dagger$ \underline{\textbf{Underlined}} best result on noisy speech mixtures (mixboth)}
       \item \footnotesize{$\ast$ \underline{\textbf{Underlined}} best result on clean speech mixtures (mixclean)}
     \end{tablenotes}
\end{threeparttable}
}
\vspace{-15px}
\end{center}
\end{table*}
\section{Main Results}
\label{sec:results}
Tables \ref{table:twospeaker}, \ref{table:der},  and \ref{table:separation} show the performance of UME compared with previous works on downstream single task frameworks. Moreover, we also compare our results and report the findings by explicitly setting the multi-task learning weights of the individual tasks to zero in our unified framework for an unbiased comparison, providing more insights about the flexibility of our proposed method. In the following sections, we discuss the experimental results in detail.

\subsection{Multi-speaker ASR results}
\label{subsec:twospeakerASR}
For the multi-speaker ASR task, we first input the OWSMv3.1 extracted features through a shallow speaker-differentiating encoder trained with CTC, attention, and PIT losses without using the SD and SS tasks ($\lambda_{\text{asr}}$=1.0). Similar to a previous study \cite{Chang} which is our reproducible baseline, we initialized the SpeakerEnc$_{\text{SD}}^{c}$ blocks (see Section \ref{subsec:multiasr}) with a pre-trained model from the ESPnet recipe for training stability. For the multi-speaker ASR task in UME, we observe that the initialization of the ASR model provides training stability and outperforms the strong baselines in Table \ref{table:twospeaker} both for 100\% overlap (Libri2Mix \& Libri3Mix) and partial overlap task (LibriSpeech2Mix \& LibriSpeech3Mix). On the ``mixclean" evaluation sets (Libri2Mix and Libri3Mix), UME with RWSE achieves a WER of 6.4\% and 15.9\%, respectively. Furthermore, our evaluation of the ``mixboth" evaluation set validates the effectiveness of the proposed method in noisy conditions. Our experiments also indicate that initializing the three-speaker model with a pre-trained two-speaker model is essential, as training without such initialization consistently resulted in divergence. Notably, the UME framework was trained using the ``mixboth'' Libri2Mix and Libri3Mix training set, which combines two-speaker and three-speaker mixtures with WHAM! noise in 100\% overlap setting but also evaluated on the LibriSpeech2Mix and LibriSpeech3Mix evaluation set containing only clean speech with partial overlap. This demonstrates its superior generalization ability across datasets with varying data modeling characteristics.

\subsection{End-to-end speaker diarization results}
\label{subsec:diarization}
Table \ref{table:der} presents the results for the UME in the SD task, which outperforms WavLM \cite{WavLM}, achieving a DER of 1.37\% in a 100\% overlapped task setting for Libri2Mix. Furthermore, UME also achieved state-of-the-art results on Libri3Mix. Notably, WavLM is trained using overlapped speech mixtures, whereas OWSMv3.1 \cite{OWSMv3.1} is trained solely on clean speech. Despite this, OWSMv3.1, adapted as the multi-speaker encoder having an improved architecture, outperforms WavLM. We believe that the additional training losses from SS and multi-speaker ASR tasks provide additional granularity during the training in the multi-task learning framework. 

\subsection{End-to-end speech separation results}
\label{subsec:separation}

\begin{figure}[!h]
    \centering
    \resizebox{0.46\textwidth}{!}{
    \begin{minipage}{\textwidth}
    \centering
    \begin{subfigure}[b]{0.50\textwidth}
        \centering
        \includegraphics[width=1.\linewidth]{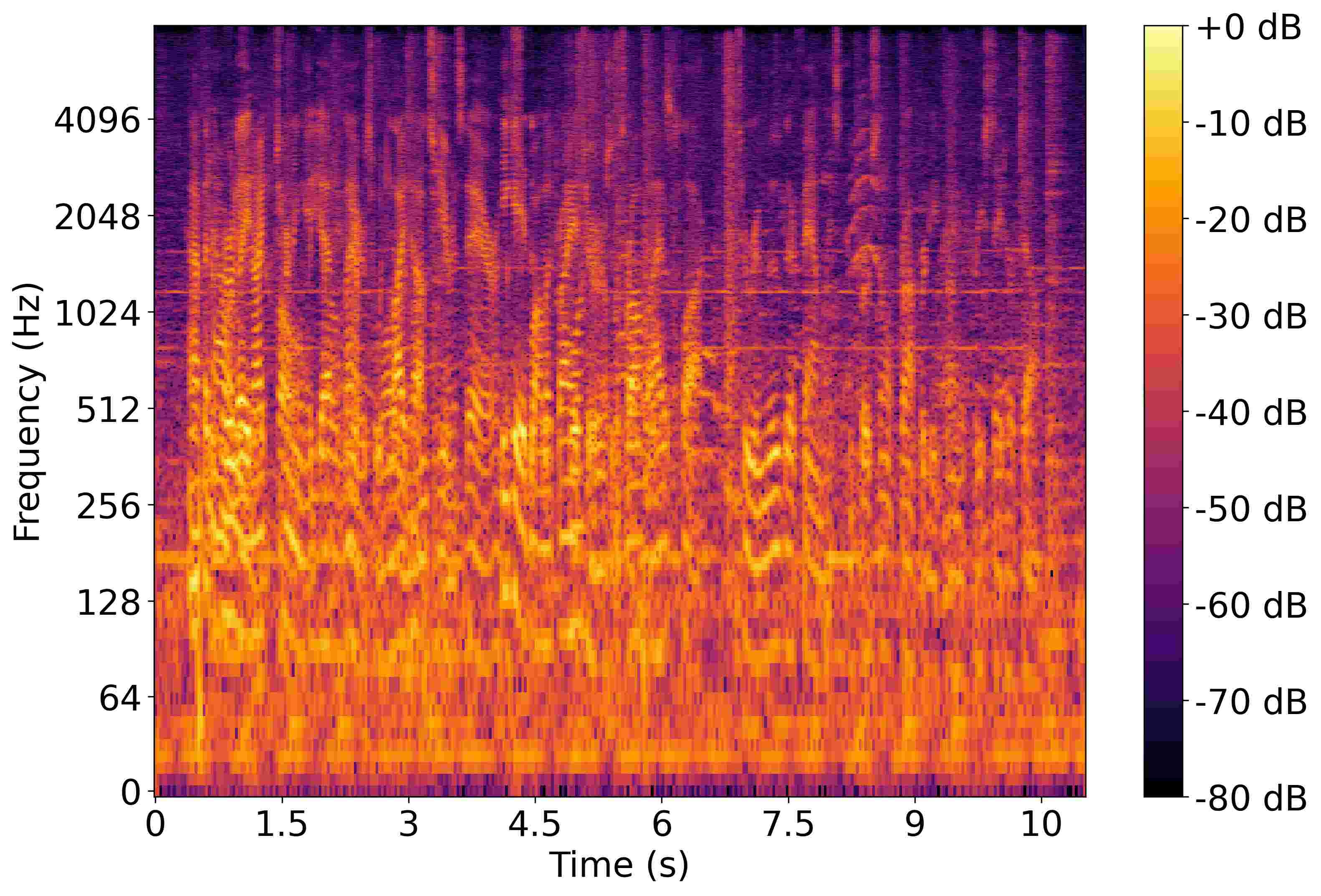}
        \caption{Input speech mixture (speaker1 + speaker2 + noise)}
    \end{subfigure}%
    \newline
    \begin{subfigure}[b]{0.50\textwidth}
        \centering
        \includegraphics[width=1.\linewidth]{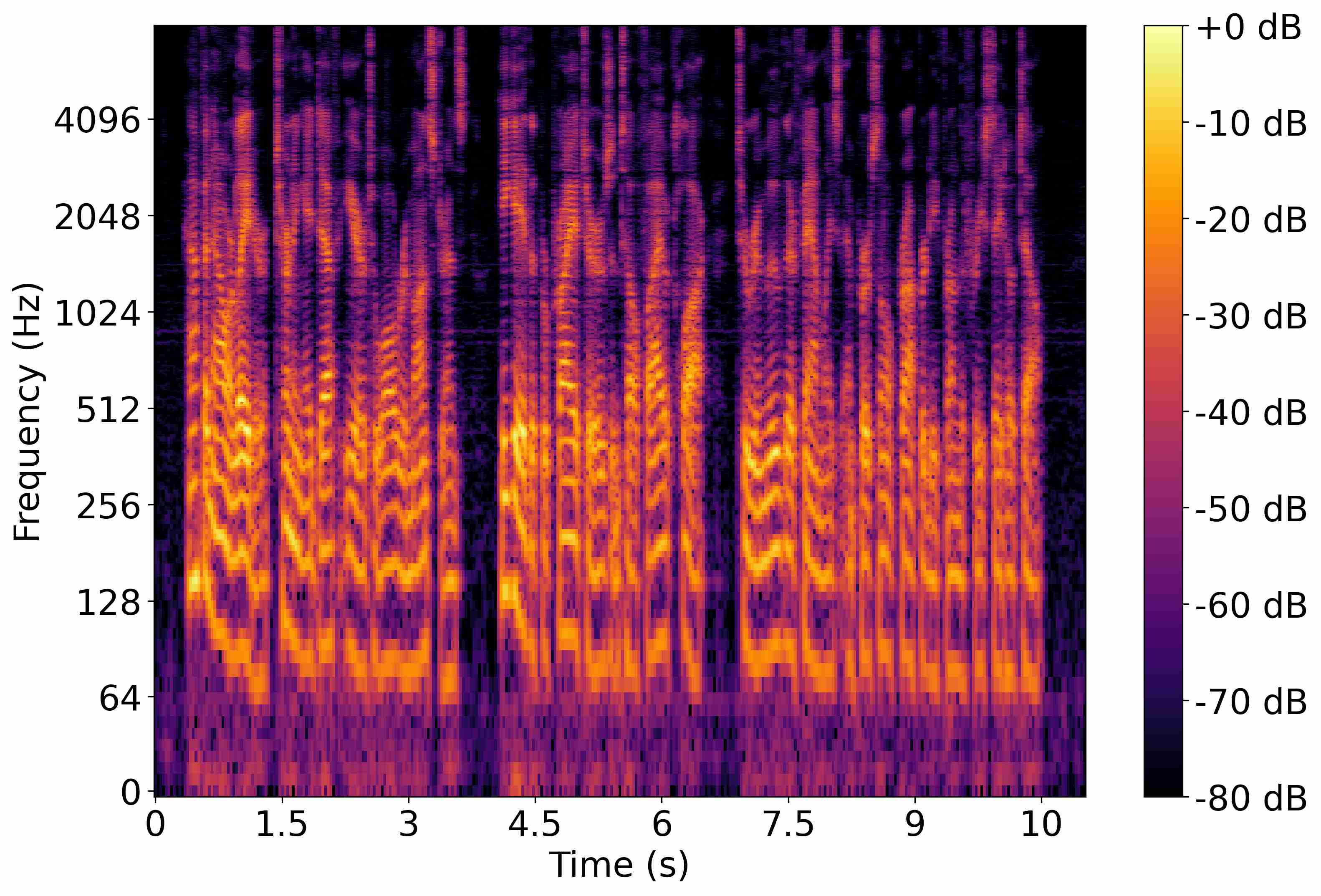}
        \caption{Ground truth speaker1}
    \end{subfigure}%
    ~
    \begin{subfigure}[b]{0.50\textwidth}
        \centering
        \includegraphics[width=1.\linewidth]{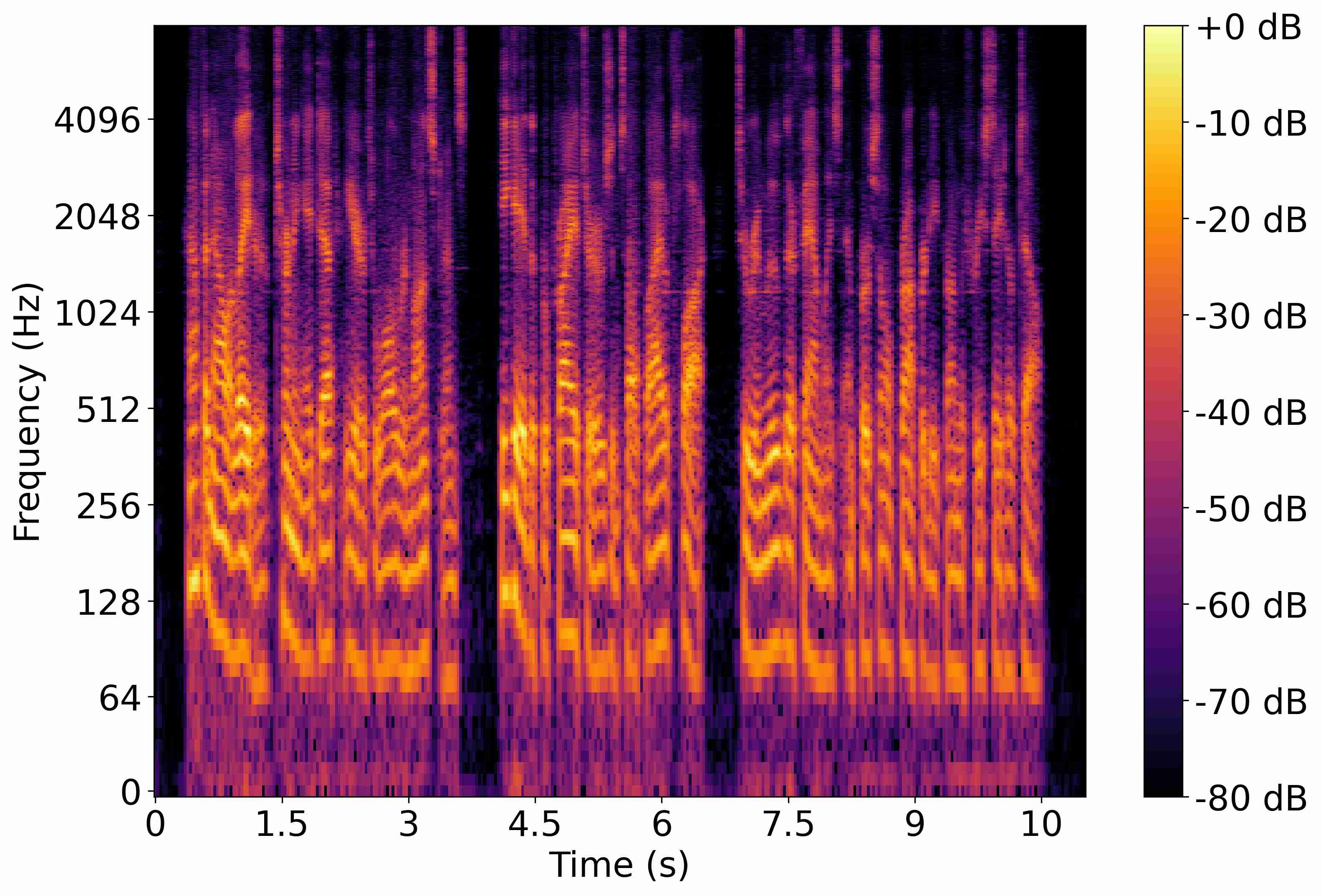}
        \caption{Recovered speaker1}
    \end{subfigure}%
    \newline
    \begin{subfigure}[b]{0.50\textwidth}
        \centering
        \includegraphics[width=1.\linewidth]{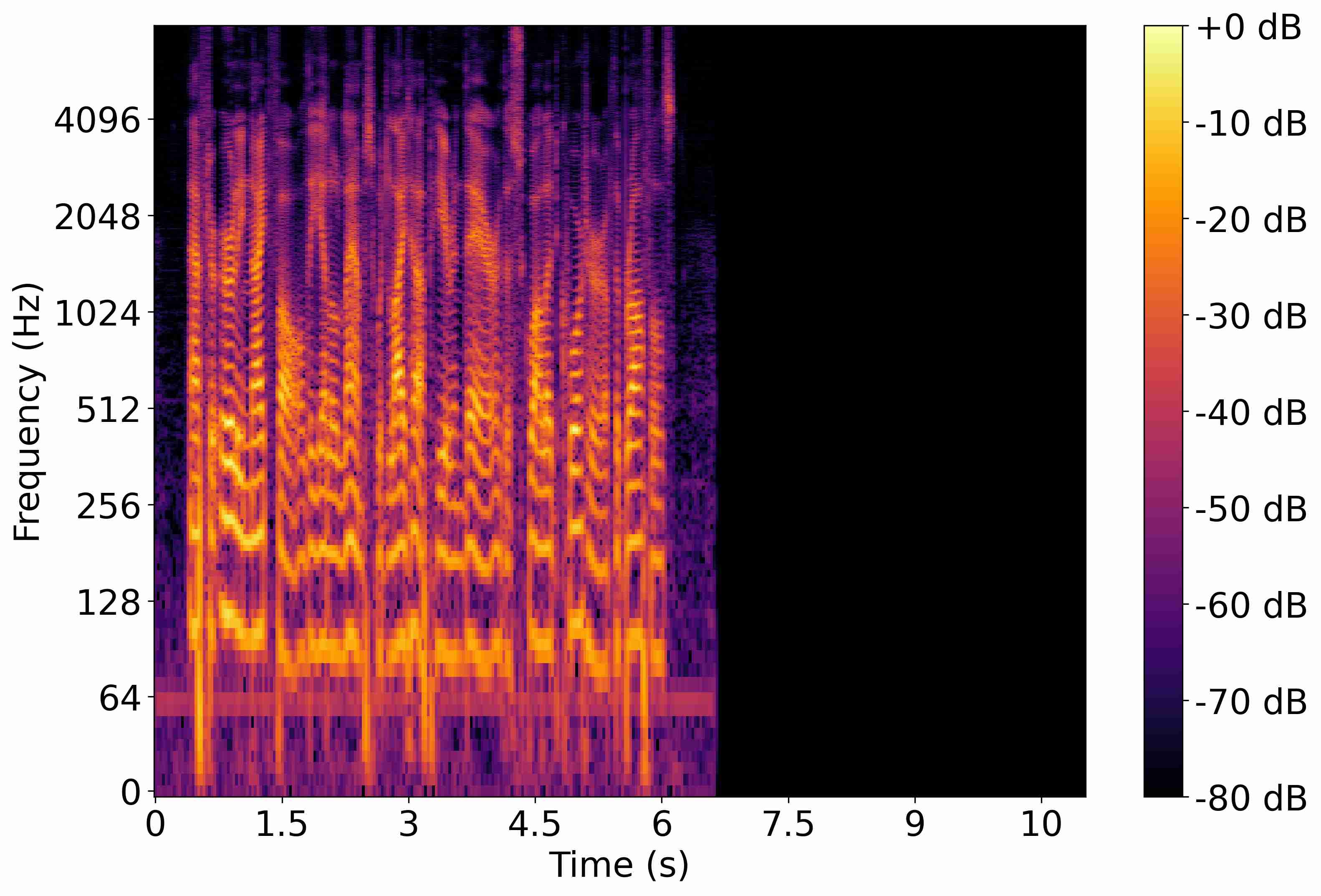}
        \caption{Ground truth speaker2}
    \end{subfigure}%
    ~
    \begin{subfigure}[b]{0.50\textwidth}
        \centering
        \includegraphics[width=1.\linewidth]{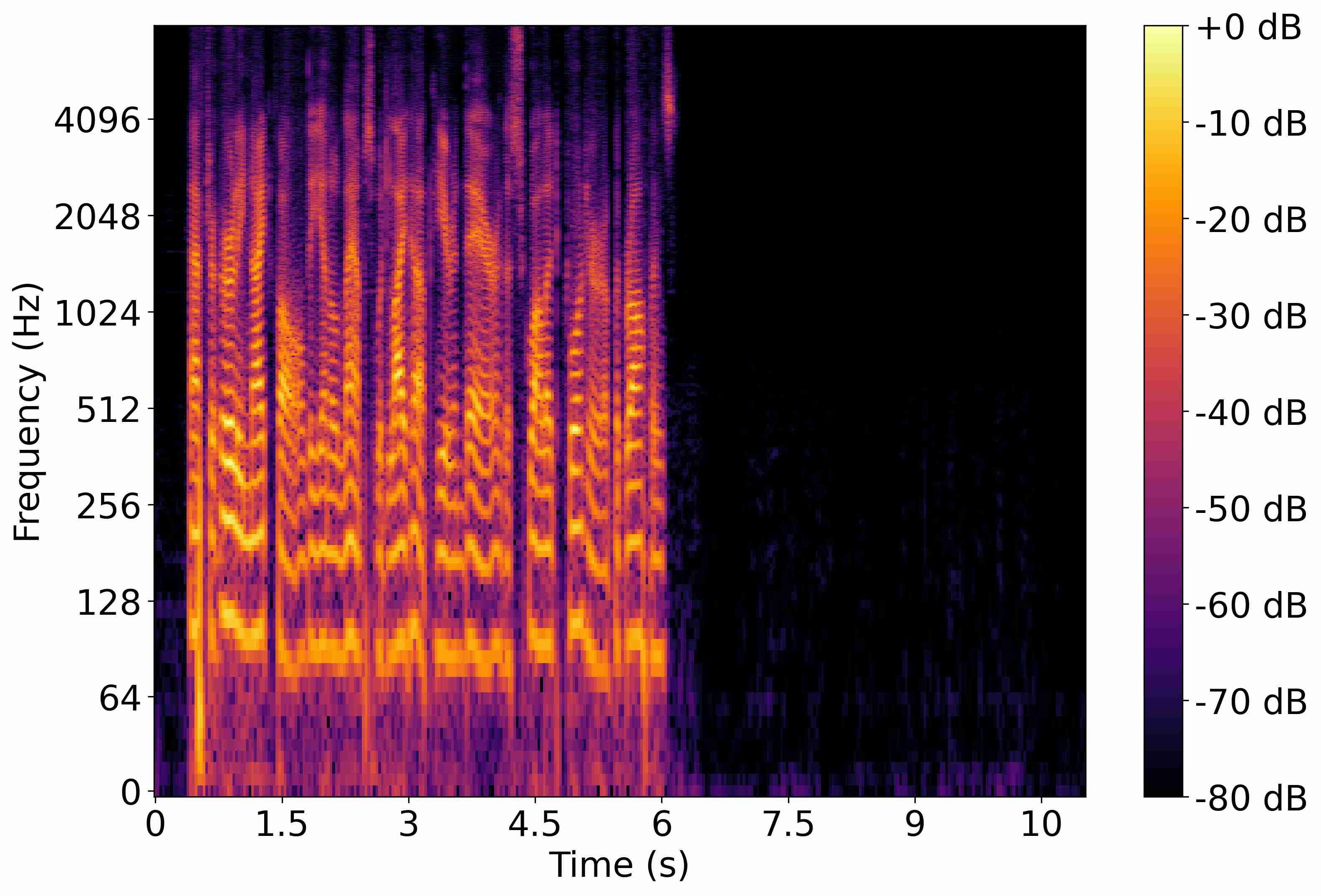}
        \caption{Recovered speaker2}
    \end{subfigure}%
    \end{minipage}
    }
    \newline
    \caption{Separation results of two speaker mixtures. (a) Input speech mixture of two speakers and WHAM! noise (speaker1, speaker2 and noise) with 100\% overlap. (column 1) Ground truth for separated signals. (column 2) Recovered speech signals using separation branch output (after concatenation).}
    \label{fig:twospeakerseparation}
\end{figure}

\begin{figure*}[!ht]
    \centering
    \resizebox{\textwidth}{!}{
    \begin{minipage}{\textwidth}
        \centering
        \begin{minipage}[c]{0.32\textwidth}
            \centering
            \includegraphics[width=\linewidth]{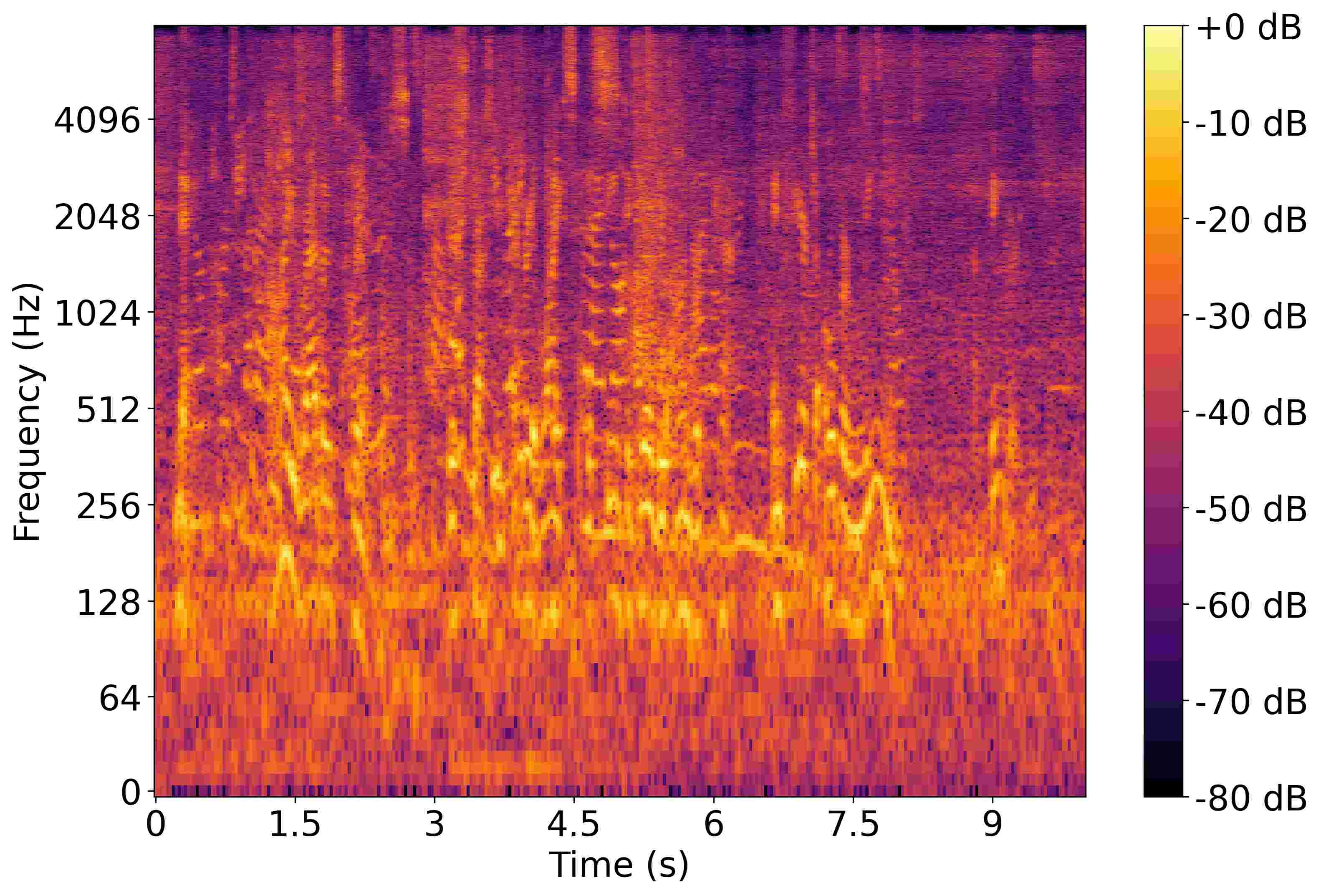}
            \caption*{(a) Input speech mixture}
        \end{minipage}
        \hfill
        \begin{minipage}[c]{0.66\textwidth}
            \centering
            \begin{minipage}[b]{0.32\textwidth}
                \centering
                \includegraphics[width=\linewidth]{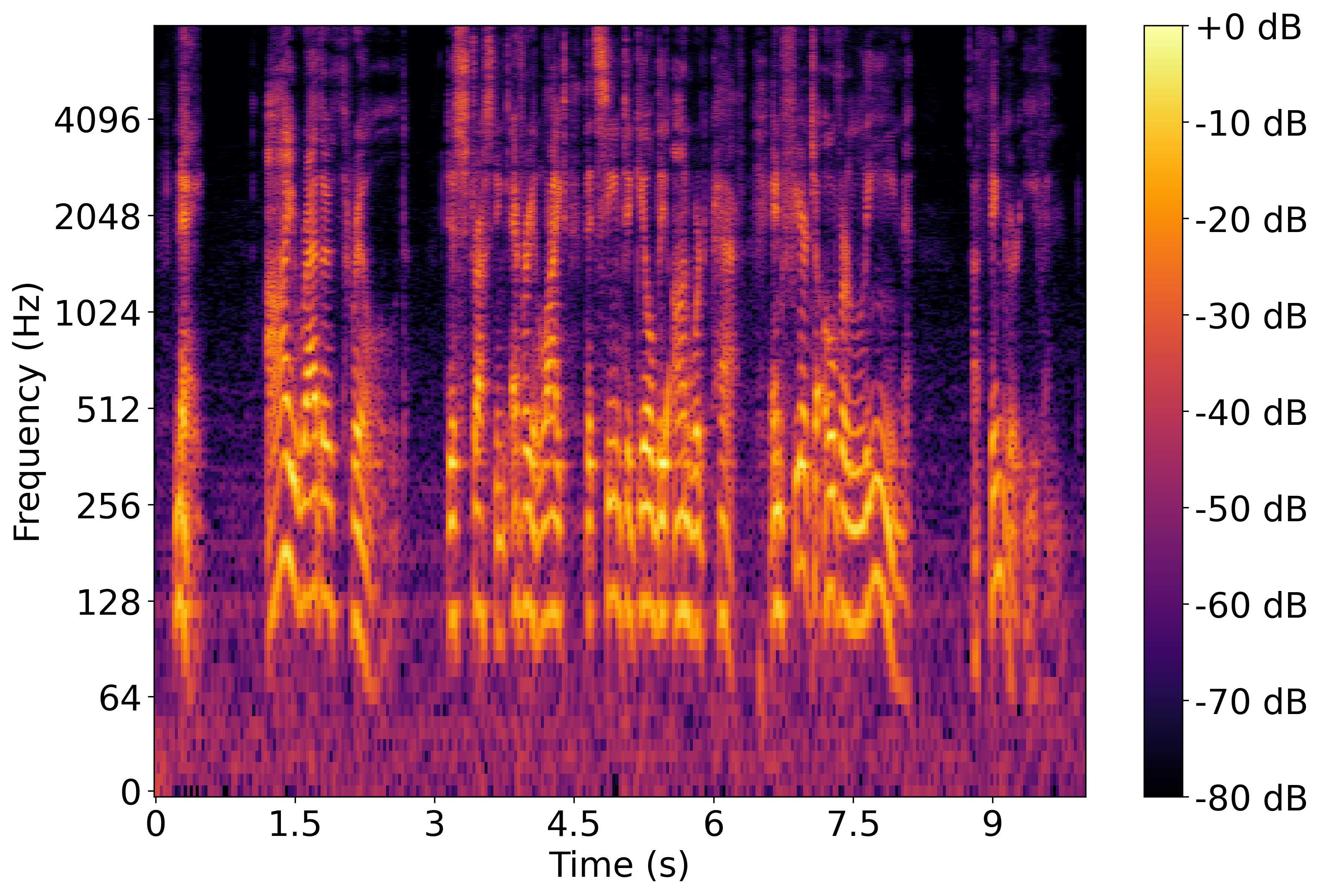}
                \caption*{(b) Ground truth speaker1}
            \end{minipage}%
            \hfill
            \begin{minipage}[b]{0.32\textwidth}
                \centering
                \includegraphics[width=\linewidth]{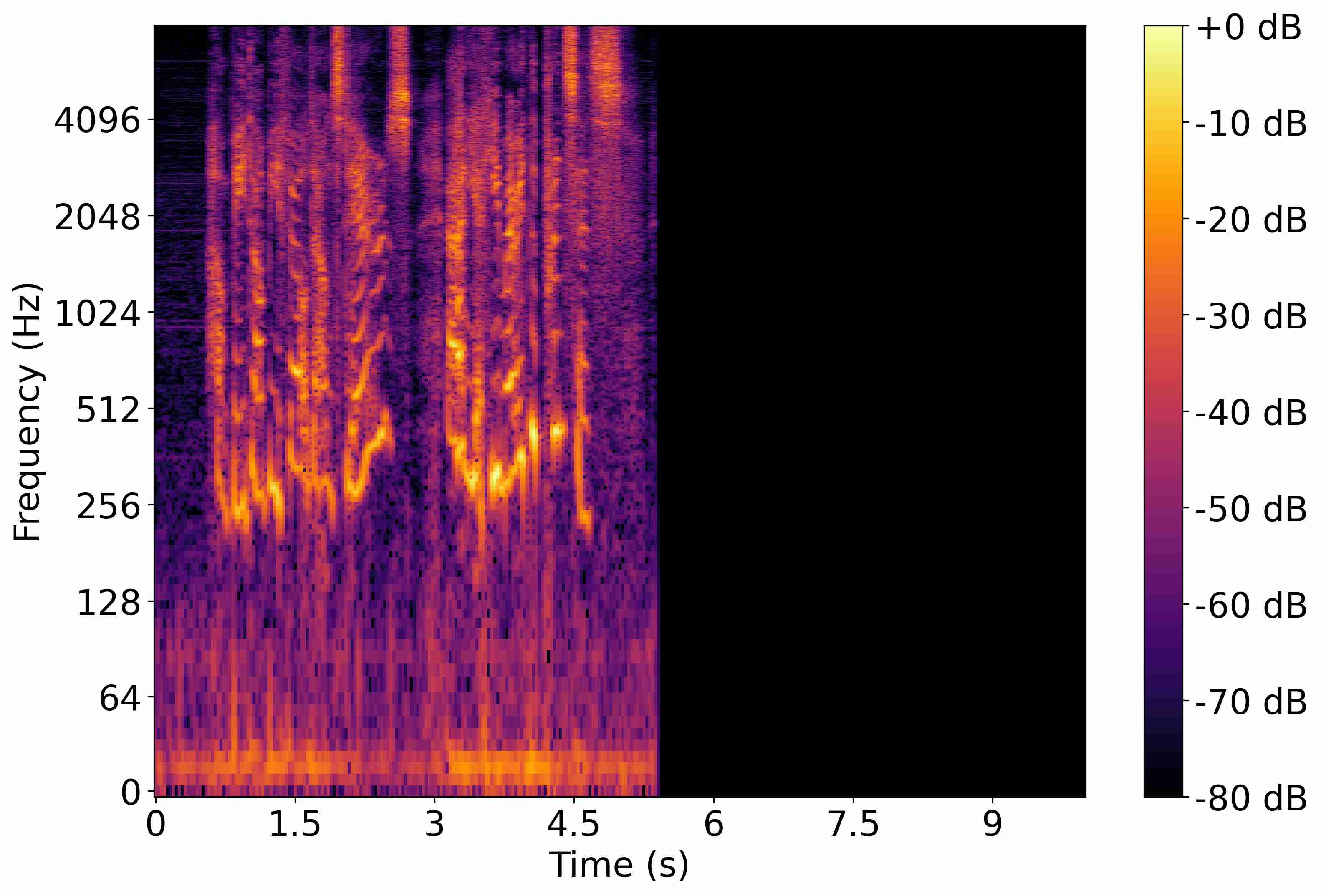}
                \caption*{(c) Ground truth speaker2}
            \end{minipage}%
            \hfill
            \begin{minipage}[b]{0.32\textwidth}
                \centering
                \includegraphics[width=\linewidth]{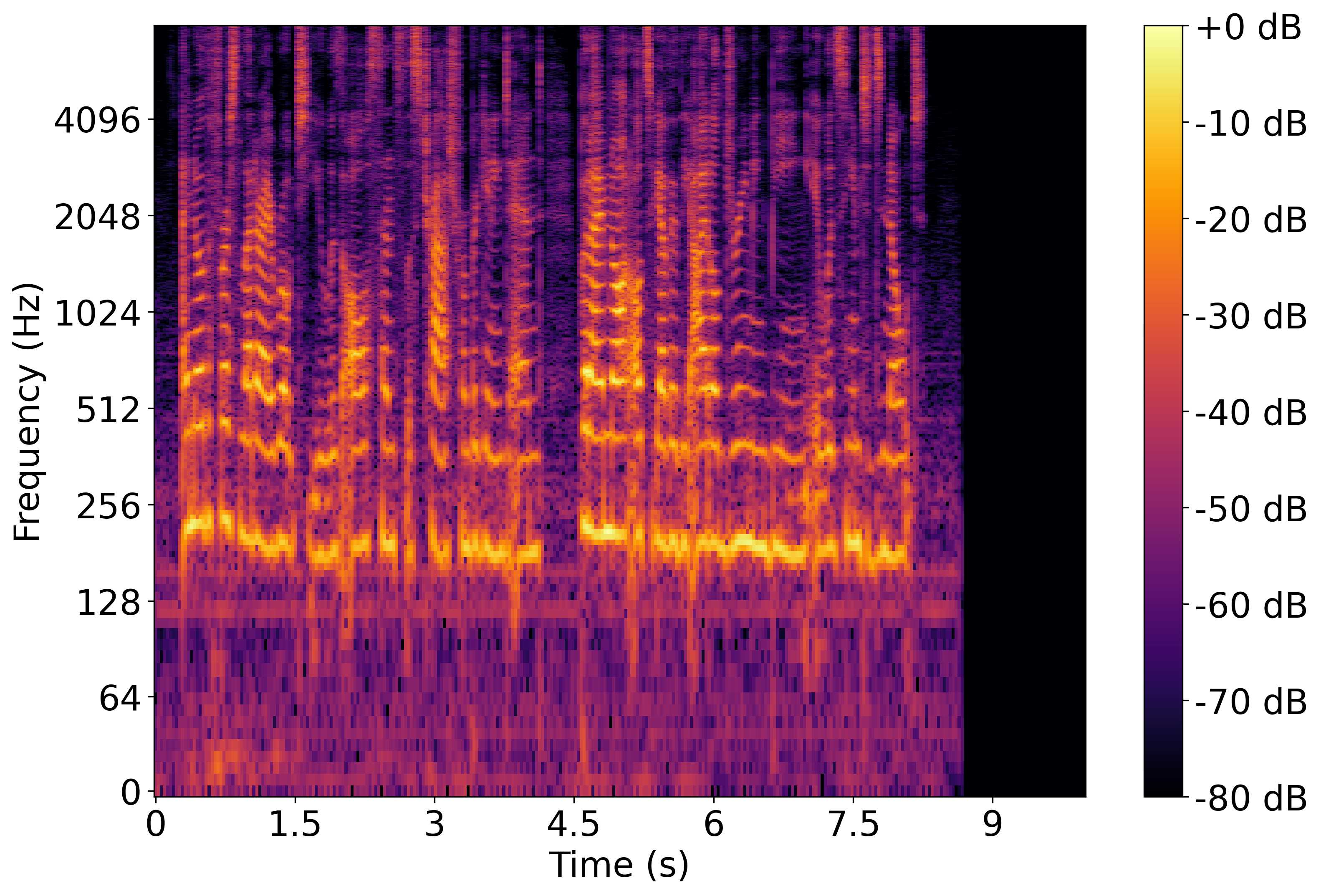}
                \caption*{(d) Ground truth speaker3}
            \end{minipage}
            \vspace{0.5em}
            \\
            \begin{minipage}[b]{0.32\textwidth}
                \centering
                \includegraphics[width=\linewidth]{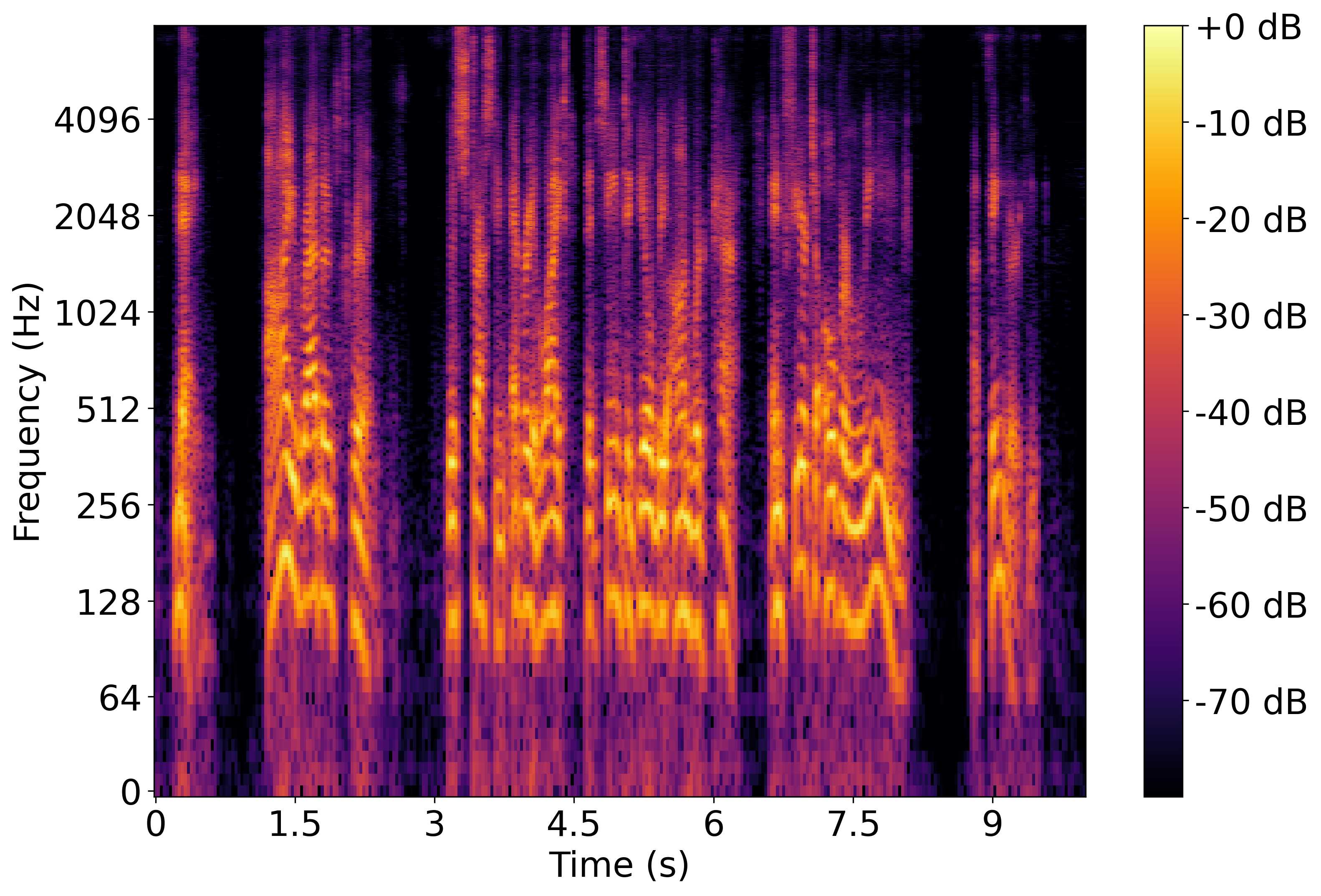}
                \caption*{(e) Recovered speaker1}
            \end{minipage}%
            \hfill
            \begin{minipage}[b]{0.32\textwidth}
                \centering
                \includegraphics[width=\linewidth]{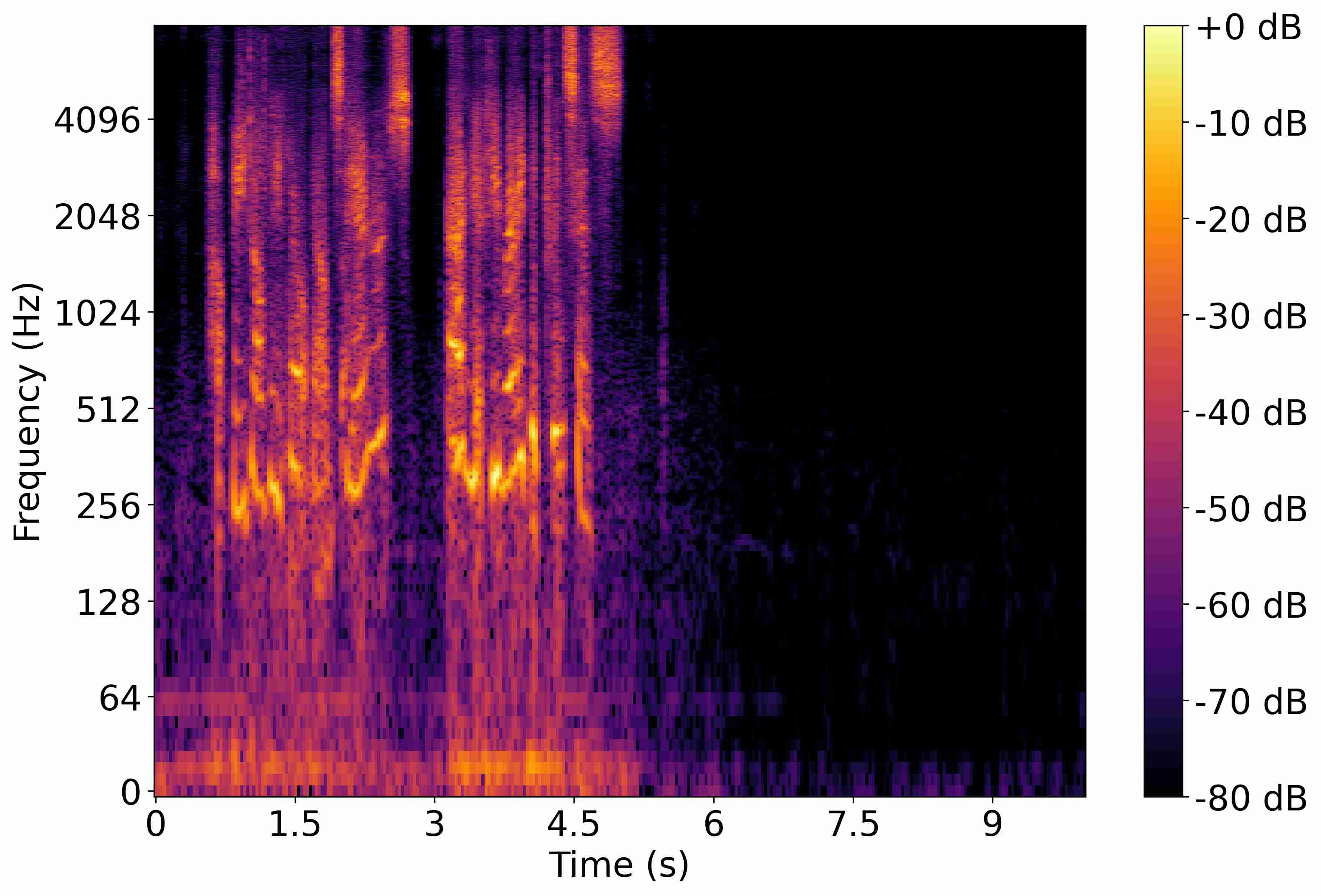}
                \caption*{(f) Recovered speaker2}
            \end{minipage}%
            \hfill
            \begin{minipage}[b]{0.32\textwidth}
                \centering
                \includegraphics[width=\linewidth]{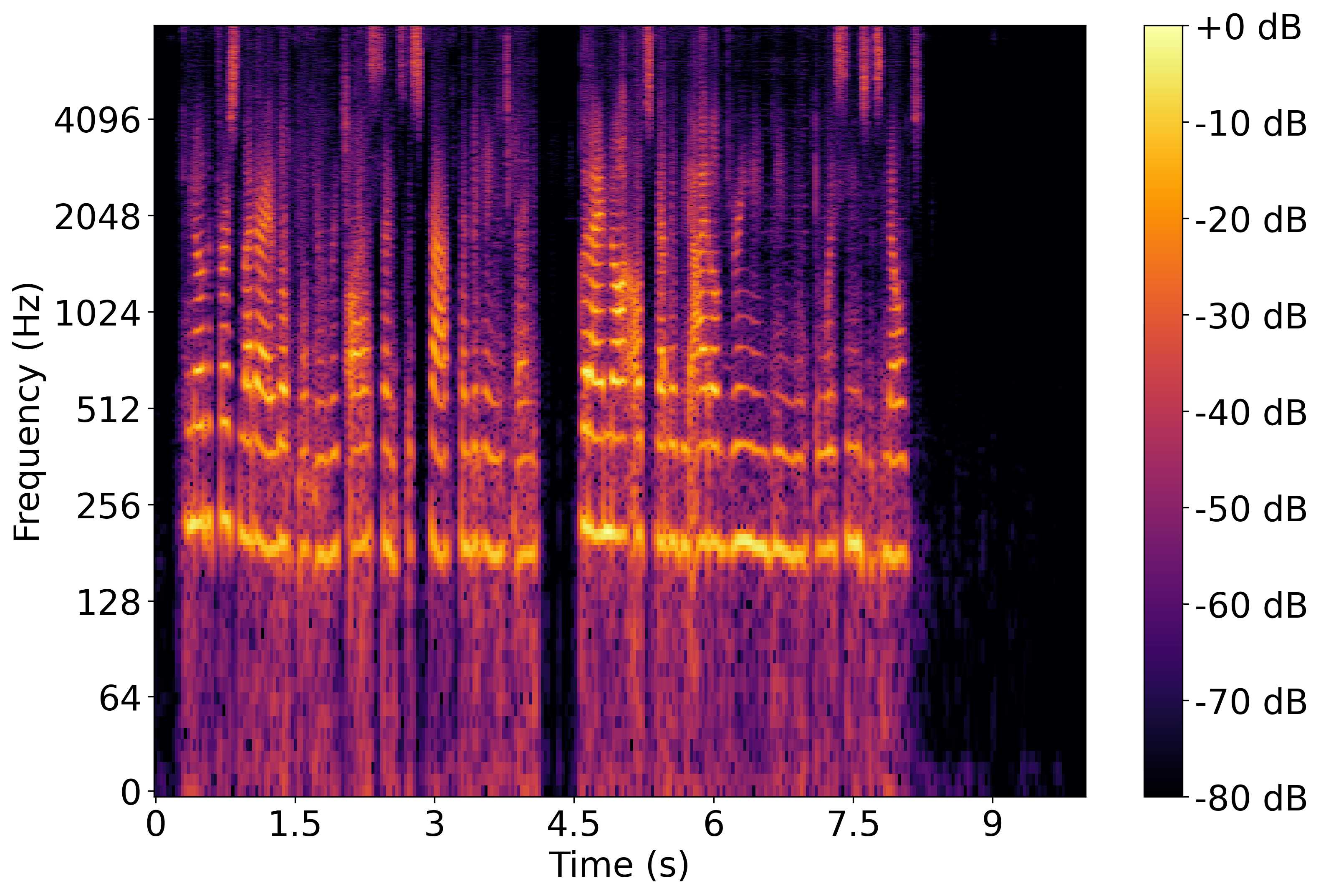}
                \caption*{(g) Recovered speaker3}
            \end{minipage}
        \end{minipage}
    \end{minipage}
    }
    \caption{Separation results of three speaker mixtures. (a) Input speech mixture of three speakers and WHAM! noise (speaker1, speaker2, speaker3 and noise) with 100\% overlap (b--d) Ground truth speech signals. (e--g) Recovered speech signals.}
    \label{fig:threespeakerseparation}
\end{figure*}

Unlike previous studies \cite{SUPERB,WavLM} which report the separation results in ``min mode'', UME requires overlapped mixtures in ``max mode'' during the training process due to the unification of the ASR task, as discussed in Section \ref{subsec:dataset}. For this reason, we evaluate the UME on 100\% overlapped mixtures in ``max mode'' with our fully reproduced Conv-TasNet model following the similar setup in Section \ref{subsec:implementation} without the concatenated features. Experimental results in Table \ref{table:separation} demonstrate consistent improvement compared to the separation-only tasks, indicating that concatenating encoded features with the upsampled hidden representations of the OWSMv3.1 encoder in the TCN block (see Section \ref{subsec:ss}) improves separation performance (see Figures \ref{fig:twospeakerseparation} \& \ref{fig:threespeakerseparation}), resulting in an improved performance in recovering the speaker activity without using an additional diarization branch.

\begin{table*}[!ht]
\caption{Speech separation results on the evaluation sets of Libri2Mix and Libri3Mix. The metrics STOI, SDR, and SI-SNR are reported in decibels (dB). ($\lambda_{\text{asr}},\lambda_{\text{diar}}, \lambda_{\text{sep}}$) denote training weights next to UME. \textbf{Bold:} the best result on noisy speech mixtures. \textbf{\underline{Underlined:}} the best result on clean speech mixtures. }
\label{table:separation}
\begin{center}
\resizebox {0.75\linewidth} {!} {
\begin{threeparttable}[b]
\begin{tabular}{@{}lllll}
    \toprule
    \multirow{2}{*}{\bf ID}                                & \multirow{2}{*}{\hspace{2em} \bf Model}                         & \multicolumn{3}{c}{\textbf{Libri2Mix / Libri3Mix}} \\
    \cmidrule(l){3-5} 
     &                                                                                                          &  \multicolumn{1}{c}{ \bf STOI}   & \multicolumn{1}{c}{ \bf SDR}     & \multicolumn{1}{c}{ \bf SI-SNR}    \\
    \midrule
    \rowcolor{gray!30} \multicolumn{5}{c}{\bf{\textit{Baseline}} $\rightarrow$ \textbf{\textit{Training set: LibriMix (460 hours, mixboth: speech $+$ noise, mode= max)}}} \\ 
    \rowcolor{gray!10}\multicolumn{1}{l}{}                &\hspace{2em} ConvTasNet \cite{Luo} (reprod.)         &  87.63 / N/A                     & 11.48 / N/A                      & 10.93 / N/A                         \\
    \midrule
    \rowcolor{pink!100} \multicolumn{5}{c}{\bf{\textit{Proposed}} $\rightarrow$ \textbf{\textit{Training set: LibriMix (460 hours, mixboth: speech $+$ noise, mode= max)}}}\\    
    \rowcolor{pink!70}\multicolumn{1}{l}{\textbf{A1}}     &\hspace{1em} \textbf{\textit{w/o weighted sum}}      &                                  &                                  &                                     \\
    \rowcolor{pink!30}\multicolumn{1}{l}{}                &\hspace{2em} UME ($\lambda_{\text{sep}} = 1.0$)      &  89.13/85.31                     & 12.39/10.16                      & 11.81/9.53                          \\
    \rowcolor{pink!30}\multicolumn{1}{l}{}                &\hspace{2em} UME (0.1, 0.1, 0.8)                     &  90.49/div.\tnote{$\S$}          & 13.18/div.\tnote{$\S$}           & 12.64/div.\tnote{$\S$}              \\
    \rowcolor{pink!70}\multicolumn{1}{l}{\textbf{A2}}     &\hspace{1em}\textbf{\textit{w/ weighted sum}}        &                                  &                                  &                                     \\
    \rowcolor{pink!30}\multicolumn{1}{l}{}                &\hspace{2em} UME (0.33, 0.33, 0.34)                  &  90.29/div.\tnote{$\S$}          & 13.05/div.\tnote{$\S$}           & 12.51/div.\tnote{$\S$}              \\
    \rowcolor{pink!30}\multicolumn{1}{l}{}                &\hspace{3em} + ASR init.                             &  89.82/86.48                     & 12.68/10.69                      & 12.12/10.07                         \\
    \rowcolor{pink!30}\multicolumn{1}{l}{}                &\hspace{2em} UME (0.1, 0.1, 0.8)                     &  \textbf{90.82}/div.\tnote{$\S$} & \textbf{13.39}/div.\tnote{$\S$}  & \textbf{12.84}/div.\tnote{$\S$}     \\
    \rowcolor{pink!70}\multicolumn{1}{l}{\textbf{A3}}     &\hspace{1em}\textbf{\textit{w/ RWSE}}                &                                  &                                  &                                     \\
    \rowcolor{pink!30}\multicolumn{1}{l}{}                &\hspace{2em} UME ASR init. (0.33, 0.33, 0.34)        &  89.95/\textbf{86.71}            & 12.76/\textbf{10.79}             & 12.22/\textbf{10.18}                \\
    \midrule
    \rowcolor{pink!100} \multicolumn{5}{c}{\bf{\textit{Proposed}} $\rightarrow$ \textbf{\textit{Training set: LibriMix (460 hours, mixclean: speech $+$ noise, mode= max)}}}                                                \\
    \rowcolor{pink!70}\multicolumn{1}{l}{\textbf{A4}}     &\hspace{1em}\textbf{\textit{w/ RWSE}}                &                                  &                                  &                                     \\
    \rowcolor{pink!30}\multicolumn{1}{l}{}                &\hspace{2em} UME ASR init. (0.33, 0.33, 0.34)        & \underline{\textbf{95.64}}/\underline{\textbf{91.25}} & \underline{\textbf{17.41}}/\underline{\textbf{13.07}} & \underline{\textbf{17.06}}/\underline{\textbf{12.58}}    \\
    \bottomrule
\end{tabular}
     \begin{tablenotes}
       \item \footnotesize{$\S$ For the three-speaker case, training diverged (div.) w/o ASR initialization.}
     \end{tablenotes}
\end{threeparttable}
}
\end{center}
\end{table*}

\section{Conclusion}
In this paper, we propose UME, a unified framework for end-to-end speech processing, which integrates speaker diarization, speech separation, and multi-speaker ASR with a residual weighted-sum encoding (RWSE) of the intermediate encoder layers. UME substantially outperforms strong baselines and previous works and achieves state-of-the-art performance on the speaker diarization task. In the future, we plan to apply this framework to more challenging multi-speaker scenarios like CHiME-6 \cite{ShinjiCHiME6}. 
We are also interested in extending it to a multilingual UME.


\bibliographystyle{IEEEtran}
\bibliography{asru2025}

\end{document}